\documentclass[aps,prl,preprintnumbers,amsmath,amssymb,latexsym,nofootinbib,array,enumerate,letter,twocolumn,superscriptaddress,longbibliography]{revtex4-2}

\usepackage{graphicx}
\usepackage[pdftex]{hyperref}
\usepackage[svgnames]{xcolor}
\definecolor{phthaloblue}{rgb}{0.0, 0.06, 0.54}
\hypersetup{
    colorlinks=true,
    linkcolor=phthaloblue,
    citecolor=blue,
    filecolor=blue,
    urlcolor=phthaloblue,
    }
\usepackage{amsmath}
\usepackage{amssymb}

\newcommand{\micromegas}{\code{MicrOMEGAs5.0}~}
\newcommand{\feynrules}{\code{FeynRules}~}
\newcommand{\code}[1]{\texttt{#1}}
\newcommand{\pr}[1]{\left(#1\right)}
\newcommand{\degree}{\ensuremath{^\circ}}
\newcommand{\imgscaling}{0.9}

\begin{document}

\title{Neutrinos for TeV Neutralinos}
\author{Carsten Rott}
\affiliation{University of Utah Department of Physics and Astronomy}
\author{Pearl Sandick}
\affiliation{University of Utah Department of Physics and Astronomy}
\author{Ben Sheff}
\affiliation{University of Utah Department of Physics and Astronomy}
\date{\today}

\begin{abstract}    
One of the most well-founded candidates for dark matter remains a split-SUSY model with a Higgsino-like or Wino-like lightest superpartner and the grander SUSY model providing answers for the hierarchy problem and GUT scale unification.The relatively heavy scalar superpartners imply such models would not yet be seen at collider experiments, and mixing-suppressed couplings place such models outside the reach of current direct detection experiments. As such particles annihilate fairly readily to electroweak bosons, a significant neutrino signal can arise near the galactic center that may be visible to dedicated searches at current and future neutrino telescopes.
\end{abstract}

\maketitle

\section{Introduction}
In the wake of the non-discovery of supersymmetry (SUSY) at the Large Hadron Collider, SUSY models with light scalars have been mostly ruled out. There remain, however, a class of models with light neutralinos and heavy scalar superpartners unreachable by modern, terrestrial experiments, commonly called split SUSY\footnote{In this paper we assume the scalars are heaver than $\mathcal{O}(100)$ TeV covering most mini-split-SUSY and split-SUSY models (the cutoff between the two being at roughly $10^5$ TeV~\cite{Arvanitaki:2012ps}), so we will collectively refer to these as split SUSY models.}. 
Split SUSY models were first proposed in 2003, and the theoretical foundations have since been explored, expanding on how such theories arise and their implications for the Higgs boson mass~\cite{wells_implications_2003,Pierce:2004mk,wells_pev-scale_2005,arkani-hamed_well-tempered_2006,Co:2022jsn,Evans:2022gom}.
Such models are attractive as they unify the gauge forces at high energies, reduce the Higgs hierarchy problem, and the lightest neutralino can serve as the dark matter (DM) in the universe. 

The primary methods to detect particle DM are production, direct detection, and indirect detection.
Production is explored through colliders, where there is sufficient energy in particle collisions to generate the DM particles directly. While the gauginos and Higgsinos may be light enough in Split SUSY models to be produced in a modern collider, the high masses and weak couplings can lead to signatures current experiments cannot detect~\cite{Baer:2011ec, ATLAS:2015wrn,Baer:2020kwz}. Direct detection is a technique in which DM particles directly scatter off of nucleons in a terrestrial detector, which can be observed by signals of nuclear or electron recoils in the detector material.  This technique typically has limited reach for high mass DM candidates.  For DM in split-SUSY models, the reach may be even further suppressed~\cite{Beylin:2007yt,xenon_1T_original,Co:2021ion,Evans:2022gom}. 

Here, we focus on indirect detection, a technique aiming to observe the products of DM annihilation which may be taking place anywhere in the universe.  Telescopes around the world and in space seek to detect annihilation products such as photons, neutrinos, or anti-particles. In particular, we focus here on the prospects for detecting neutrinos from neutralino annihilation near the galactic center (GC) where the density of DM is expected to be high. Similar studies have been conducted in the past regarding anti-particles~\cite{Hisano:2005ec, Hryczuk:2014hpa}, and  photons~\cite{Vereshkov:2005ad,Beylin:2007yt,Cohen_wino_2013,Fan:2013faa,slatyer_prospects_2021,Dessert:2022evk}, which both serve as complementary channels, with different systematics and backgrounds, to our analysis of the sensitivities of modern neutrino experiments.

To demonstrate the current and near future experimental sensitivities to neutrinos produced from these neutralino annihilations, we will focus on IceCube~\cite{IceCube:2016zyt}, the largest current neutrino detector, and the Cubic Kilometer Neutrino Telescope (KM3NeT)~\cite{KM3Net:2016zxf}, which is expected to complete construction in 2030~\cite{Biagi:2023yoy}. As we demonstrate here, IceCube has significant capability to observe Wino-like DM, and is expected to increase that reach across model space with the IceCube Gen2 upgrade~\cite{Ishihara:2019aao,Clark:2021fkg}. We also compare the IceCube sensitivity to the projected sensitivity for KM3NeT.

We first discuss the details of dark matter annihilation in split-SUSY models in Section~\ref{sec:DMAinSS}, in particular a discussion of the nature of the DM as a mixed state dominated by either the Higgs or W-boson superpartners, and enhancements from Sommerfeld effects. In Section~\ref{sec:IDwN}, we will walk through the calculation of astrophysical effects and of the spectrum of neutrino flux expected from DM annihilation. We will then discuss the particular detector signatures expected in IceCube and KM3NeT from such a signal and the calculation of their respective sensitivities. Finally, in Section~\ref{sec:Res}, we will project the sensitivities of both of these experiments in dedicated searches for neutralino DM, and conclude with Section~\ref{sec:Con}.

\section{Dark Matter Annihilation in Split-SUSY}
\label{sec:DMAinSS}
In split-SUSY models, the scalar superpartners are heavy enough to effectively decouple, while gauginos and Higgsinos are comparatively light. There are a number of theories that can lead to this structure, for example a unified SUSY-breaking mass for the scalars with conformal anomaly-induced masses for the gauginos results in a light wino $\sim 300\times$ lighter than the scalars~\cite{hitoshi_gaugino_anomaly_1998, randall_out_1999,Gherghetta:1999sw}. The accessible new physics is determined by the four neutral and two charged beyond Standard Model (BSM) fermions typical to supersymmetric extensions of the Standard Model. Specifically, the neutral particles are referred to as the Bino \smash{\big{(}$\widetilde{B}$\big{)}}, Wino \smash{\big($\widetilde{W}$\big{)}}, and down- and up-type Higgsinos \smash{\big($\widetilde{H_d}, \widetilde{H_u}$\big)}, which serve as superpartners to the hypercharge and weak gauge bosons and the Higgs bosons, respectively. Collectively these are referred to as the neutralinos, as their gauge eigenstates mix through electroweak (EW) symmetry breaking. In particular, for neutralino states arrayed in the order \smash{\big($\widetilde{B}, \widetilde{W}, \widetilde{H_d}, \widetilde{H_u}$\big)}, the neutralino mass matrix is given by
\begin{equation}
   M_0=\left(\begin{array}{cccc}
M_{1} & 0 & -g^{\prime} v_{d} / \sqrt{2} & g^{\prime} v_{u} / \sqrt{2} \\
0 & M_{2} & g v_{d} / \sqrt{2} & -g v_{u} / \sqrt{2} \\
-g^{\prime} v_{d} / \sqrt{2} & g v_{d} / \sqrt{2} & 0 & -\mu \\
g^{\prime} v_{u} / \sqrt{2} & -g v_{u} / \sqrt{2} & -\mu & 0
\end{array}\right),
\end{equation}
where $v_u$ and $v_d$ are the up- and down-type Higgs field vacuum expectation values, respectively, and $g$ and $g'$ are the weak and hypercharge couplings. $M_1$ and $M_2$ are the SUSY-breaking gaugino masses for the Bino and Wino, respectively, and $\mu$ is the $\mu$-term in the superpotential, which in these models roughly corresponds to the Higgsino mass\footnote{For a more detailed review of these parameters and of the neutralino and chargino masses, see Ref.~\cite{martin_supersymmetry_1998}}. 

The two charged BSM fermions are the partners to the electrically-charged weak bosons and the charged components of the Higgs fields. These are labelled \smash{$\widetilde{W}^{\pm}$} and \smash{$\widetilde{H}_{u}^{+}, \widetilde{H}_{d}^{-}$}, and are collectively referred to as charginos.  The mass matrix for field order \smash{\big($\widetilde{W}^{+}, \widetilde{H}_{u}^{+}, \widetilde{W}^{-}, \widetilde{H}_{d}^{-}$\big)} can be expressed as
\begin{equation}
    \begin{gathered}
\mathbf{M}_{+}=\left(\begin{array}{cc}
\mathbf{0} & \mathbf{X}^{T} \\
\mathbf{X} & \mathbf{0}
\end{array}\right), \quad {\rm for } \quad
\mathbf{X}=\left(\begin{array}{cc}
M_{2} & g v_{u} \\
g v_{d} & \mu
\end{array}\right).
\end{gathered}
\end{equation}

As we can see, the mixing angle between the gaugino and Higgsino gauge states is of order $g v_u / |M_2 - \mu| \sim \mathcal{O}(m_Z / |M_2 - \mu|)$ for Z-boson mass $m_Z$, assuming $M_2 < M_1$.  This implies a generically small mixing angle since the energies considered below are above 300 GeV, and in many constructions of split-SUSY models $M_2$ and $\mu$ differ by at least an order of magnitude. In light of this, throughout this paper we will focus on a fairly pure Higgsino-like or Wino-like particle as the lightest superpartner (LSP), serving as a DM candidate. Each has a component of the other mixed in as well as a small bino component, which collectively induce small EW couplings that while not of consequence here, can be significant for other searches~\cite{Co:2022jsn}.

A feature of particular note in these scenarios is that reproducing the observed DM relic abundance sets fairly stringent limits on the mass of the lightest neutralino. In the case of a Higgsino-like LSP, this is $1.1\pm 0.2$ TeV~\cite{Profumo_Statistical_2004,giudice_split_2005,Pierce:2004mk,Hryczuk:2010zi}, and for a Wino-like LSP, this is $2.8\pm 0.2$ TeV~\cite{Hisano:2006nn,Cirelli:2007xd,Hryczuk:2010zi}. As long as the mixing angle is sufficiently small, the annihilation cross section for the lightest neutralino is dominated by annihilation into massive EW gauge bosons with a chargino mediator. In the case of a Wino-like LSP, the tree level masses of the lightest neutralino and lightest chargino shown above are degenerate, broken predominantly by EW radiative corrections at the one-loop level, giving a mass splitting of $\mathcal{O}(100)$ MeV~\cite{ibe_mass_2013}. In the case of a Higgsino-like LSP, the tree-level mass splitting from the matrices above is of order $m_\mathrm{Z}^2 / M_i$ where $M_i$ is the lighter of the two gaugino masses, and may be competitive with similar radiative corrections. In either case, the neutralino annihilation cross section is largely independent of the physics of other superpartners, depending solely on the lightest neutralino mass. This means the relic density makes a direct prediction for an annihilation signal, providing a definitive signal to seek with indirect detection experiments.

In order to observe neutralino-like DM, we focus in particular on neutrinos produced in  DM-DM annihilation.  The analysis pileline is the following:  We code a description of the model of interest using \feynrules~\cite{feynrules}, which generates the input for \micromegas~\cite{micromegas}, a package for calculating relic densities and direct and indirect detection signals for DM models. For the split-SUSY models studied here, non-perturbative contributions to the annihilation cross section can be large.  We use \micromegas to calculate the perturbative neutralino annihilation cross section, along with the ratio of various prompt annihilation products. These products are dominated by EW gauge bosons, with a near even split between W- and Z-bosons for Higgsino-like DM (55\% and 44\% respectively), and almost entirely W-bosons for Wino-like DM. \micromegas computations also confirm the appropriate relic density for 1.1 TeV Higgsino-like DM and 2.8 TeV Wino-like DM.

We then include the so-called Sommerfeld effect~\cite{Sommerfeld:1931qaf}, an enhancement to the DM-DM annihilation cross section due to long-range, non-perturbative interactions. The enhancements come from the effective Yukawa potential from the heavy EW gauge bosons. Such enhancements are classical effects, and are calculated by solving the Schr\"odinger equation with the relevant potential. Further details can be found in Refs.~\cite{Cohen_wino_2013, slatyer_prospects_2021}. Here, we use the numerical results from Refs.~\cite{Hisano:2004ds, Hryczuk:2011vi}; however, care was taken with the former reference to include more recent loop-level calculations for the perturbative Higgsino annihilation cross section by comparing to the \micromegas calculations with Higgsino mass far from the Sommerfeld resonance regime.

Sommerfeld enhancements are $\mathcal{O}(1)$ for Higgsino-like DM with mass near the thermal-relic-abundance expectation of $1.1$ TeV. For heavier Higgsino-like DM\footnote{Note that this could be relevant given a mechanism to suppress the DM abundance today, so as not to overclose the universe.} or Wino-like DM, the Sommerfeld enhancements form a resonance structure due to loose bound states between the DM particles. This enhancement can be as large as several orders of magnitude, as we will see below. The particular mass scale of this resonance effect is sensitive to the coupling in the annihilation process and the spectrum of particles near in mass to the DM. For neutralino mass splittings much above the weak scale, both of these are parameterized dominantly by the mass of the lightest neutralino, with subdominant effects from mixing with heavier neutralinos and charginos, expected to be $\lesssim 1$\% for TeV-scale neutralinos, and negligible radiative effects of other, higher scale SUSY parameters suppressed by loop factors and the high scalar scale.

\section{Indirect Detection with Neutrinos}
\label{sec:IDwN}

We assume the Milky Way DM density distribution follows a Navarro-Frenk-White (NFW) profile~\cite{Navarro:1995iw, Navarro:1996gj} centered at the GC. Specifically, the density profile is given by
\begin{equation}
    \rho\pr{r} = \frac{\rho_s}{\frac{r}{r_s}\pr{1 + \frac{r}{r_s}}^2}
\end{equation}
for scale radius $r_s = 24.42$ kpc and scale density $\rho_s = 0.184$ GeV/cm$^3$~\cite{Cirelli:2010xx}, roughly mapping to a density of
0.3 GeV/cm$^3$ at the solar circle. The astrophysical dependence of the neutrino (or any other indirect detection) signal is characterized by a J-factor, which for an annihilation process is
\begin{equation}
    J = \int d\Omega \int ds ~\rho^2(s, \Omega),
\end{equation}
which is integrated over the line-of-sight distance $s$ and over solid angle $\Omega$.

We use the Poor Particle Physicist's Cookbook for Dark Matter Indirect Detection (\code{PPPC4DMID})~\cite{Ciafaloni:2010ti,Cirelli:2010xx} for the neutrino spectra from neutralino annihilation, given the branching fractions to W- and Z-boson final states as discussed in Section~\ref{sec:DMAinSS}. 
We include the effects of neutrino oscillations between the production point and the detector for each neutrino flavor
using the mixing angles published by the Particle Data Group~\cite{ParticleDataGroup:2022pth}. The resulting differential neutrino flux is
\begin{equation}
    \frac{d\phi_i}{dE} = \frac{J \left<\sigma v\right>}{8\pi m_\chi^2} \sum_{\text{flavor }j} \mathcal{O}_{ij} \frac{dN_j}{dE},
    \label{eq:sigFlux}
\end{equation}
where $\left<\sigma v\right>$ is the thermally averaged velocity times cross section of the DM, $m_\chi$ is the DM mass, $\mathcal{O}_{ij}$ is the neutrino oscillation matrix\footnote{For source distances of this length, the neutrinos are fully mixed well before arrival at Earth, so the flavor ratio is fixed.} from flavor $j$ to flavor $i$, and $dN_j/dE$ is the differential expected number of neutrinos of flavor $j$ and energy $E$ produced per collision of DM.

Within a Cherenkov neutrino detector, there are typically two types of event signatures from neutrinos: tracks and cascades/showers. We focus in particular on how these are detected at IceCube, although we will compare the final sensitivity to those of the planned IceCube Gen2 and KM3NeT detectors. Tracks come primarily from muon-neutrinos ($\nu_\mu$) through charged current interactions, with sub-leading component from muon decays in taus produced in tau neutrino charged current interactions. For TeV-scale neutrinos, tracks have a median angular resolution of roughly 1\degree\cite{IceCube:2019cia}. Cascades can come from any type of neutrino with neutral current or non-$\nu_\mu$ charged current interaction.  The angular resolution
for cascades at 1 TeV is roughly 10\degree\cite{IceCube:2023ame}. In either case, higher energy neutrinos can significantly improve the possible angular resolution.

The interaction cross sections, detector volume, and effects from analysis cuts are encapsulated by a total effective area of the detector to each flavor of neutrino. This has been calculated for both tracks and cascades, seen for the Southern hemisphere for tracks in Fig.~5 of Ref.~\cite{IceCube:2019cia}, (for a neutrino of energy 1 TeV, roughly 30 cm$^2$). For cascades, the effective area is approximated by the all-flavor area in Fig.~2 of Ref.~\cite{IceCube:2023ame} (for a neutrino of energy 1 TeV, roughly 700 cm$^2$), as the electron and tau neutrino fluxes dominate in that sample. The flux times the effective area is then integrated over a bin in observed neutrino energy. For simplicity, the energy resolution is emulated through convolution with a box in log-space, assuming 15\% energy resolution\footnote{This resolution is well justified for cascades~\cite{IceCube:2017zho}, but is rather optimistic for tracks~\cite{IceCube:2021xar}}. The flux is also integrated over an angular circle about the GC taken to be 1\degree ~for tracks and 10\degree ~for cascades for reasons detailed below. The background flux is treated similarly, using astrophysical neutrinos as measured for tracks-focused analyses taken to follow a power law~\cite{IceCube:2021uhz}, with a slightly different all-flavor power law measurement used for cascades~\cite{Silva:2023wol}. Assuming Poissonian errors and a relatively small signal to background ratio, the significance of a measurement of neutrino events near the GC in an experiment is
\begin{equation}
\label{eq:significance}
\begin{split}
    \frac{S}{\sqrt{B}} =& \frac{\int_{\Delta E}dE_\nu \int_{\Delta\Omega} d\Omega ~A_{\mathrm{eff}}~ d\phi_{\nu}/dE_\nu}{\sqrt{\int_{\Delta E}dE_\nu~A_\mathrm{eff}\Delta\Omega ~ d\phi_\mathrm{astro} / dE_\nu}} ,
\end{split}
\end{equation}
where $S$ and $B$ are the expected signal and background events respectively, $\Delta E$ represents a bin in neutrino energy over which the signal and background are integrated, $\Delta \Omega$ represents the area of interest about the galactic center, $d\phi_{\nu}/dE_\nu$ is given in Eq.~\ref{eq:sigFlux}, $d\phi_{\mathrm{astro}}/dE_\nu$ is the background flux discussed above, and $A_\mathrm{eff}$ is the effective area of the detector as a function of the observed neutrino energy. We consider a single energy bin, with size and central value chosen to maximize the 
significance of a Higgsino-like, thermally produced DM indirect detection signal.  This was found to be a bin from 0.6 to 1.1 times the mass of the neutralino dark matter. The effects of changing the energy bin for different model points or of using multiple energy bins were marginal.

Note that in the calculation of the significance (Eq.~\ref{eq:significance}), the neutralino annihilation cross section appears only as a linear factor in the flux (Eq.~\ref{eq:sigFlux}). We can then factor out the annihilation cross section as $S = s\left<\sigma v\right>$ for $s$ independent of $\left<\sigma v\right>$. Given a target sensitivity to new physics at 95\% for exclusion, i.e.~a $2\sigma$ result, we have $S/\sqrt{B} = 2$, or a sensitivity to the neutralino annihilation cross section of $\left<\sigma v\right> = 2\sqrt{B}/s$.

The background sources are, in descending order of flux, atmospheric muons, atmospheric neutrinos, and astrophysical neutrinos, with at least an order of magnitude difference in each of these sources. For a source in the Southern hemisphere (as the GC is at -29\degree~declination), characteristics of the shape of the signal in IceCube help suppress down-going atmospheric muon backgrounds that aren't blocked by scattering in the Earth, while atmospheric $\nu_\mu$, the dominant contribution to atmospheric neutrinos, can be suppressed through coincidence with muons in down-going neutrino events. For cascades, this is accomplished with a deep learning algorithm, which has been used to detect the galactic plane from neutrinos at roughly 10\% of astrophysical background neutrino flux along the plane of the Milky Way~\cite{IceCube:2023ame}. For tracks, atmospheric $\nu_\mu$ rejection has been explicitly studied above 100 TeV where it can be reduced by as much as $2\times 10^{-5}$~\cite{Tosi:2019nau}. While this reference indicates that at 1 TeV this suppression is closer to a factor of 0.1, it is unclear if further improvements are possible in dedicated searches. Such additional filters will likely also reduce the expected signal, although the details will have to be explored in future analyses. 

For this study, we consider both a conservative case and an optimistic one. In the conservative case, we measure cascade events from electron and tau neutrinos (and respective anti-neutrinos) and track events from muon neutrinos and anti-neutrinos, with a background of a combination of atmospheric and astrophysical neutrinos of those types. In the optimistic case the atmospheric neutrinos are assumed to be suitably suppressed, leaving only an astrophysical neutrino background equally distributed among the flavors, which cannot be filtered as the signature is expected to be identical to that of our signal neutrinos. Atmospheric neutrino fluxes are approximated using the Honda model~\cite{Honda:2006qj}, and the astrophysical neutrino fluxes are approximated with a power law which is uniform across all flavors~\cite{Silva:2023wol}
\begin{equation}
\begin{split}
    &\frac{d\phi_{\nu, \mathrm{astro}}}{dE_\nu} = \frac{1.68^{+0.19}_{-0.22}\times 10^{-18}\times \pr{\frac{E_{\nu, \mathrm{astro}}}{100\mathrm{~TeV}}}^{-\gamma_\mathrm{astro}}}{ \mathrm{~GeV~cm^{2}~s~sr}},
    \label{eq:astro}
\end{split}
\end{equation}
for spectral index $\gamma_\mathrm{astro}=2.58^{+0.10}_{-0.09}$ and neutrino energy $E_{\nu, \mathrm{astro}}$. This law is known to be accurate at energies 3-550 TeV, and is taken as a conservative estimate in our energy range. In contrast, the most recent fully published IceCube result~\cite{IceCube:2021uhz} predicts roughly an order of magnitude smaller astrophysical flux if extrapolated to 1 TeV from its region of validity above 30 TeV, but it is two years older than the proceedings in Ref.~\cite{Silva:2023wol}. This uncertainty is encapsulated by the uncertainty on the spectral index.

\begin{figure}
    \centering
    \includegraphics[width=\imgscaling\columnwidth]{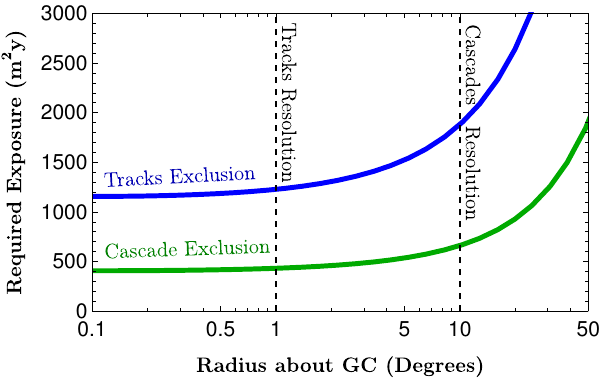}
    \caption{Total exposure for 2$\sigma$ exclusion of thermally-produced Higgsino DM as a function of the radius of the cone of interest about the GC with astrophysical neutrino background. For cascades (lower, green curve), all neutrino flavors are included for both the signal and background, while for tracks (upper, blue curve) only muon neutrinos are used. The TeV-scale angular resolution of IceCube is shown in vertical dashed lines for track and cascade signatures in the detector.}
    \label{fig:angular}
\end{figure}

\begin{figure}
    \centering
    \includegraphics[width=\imgscaling\columnwidth]{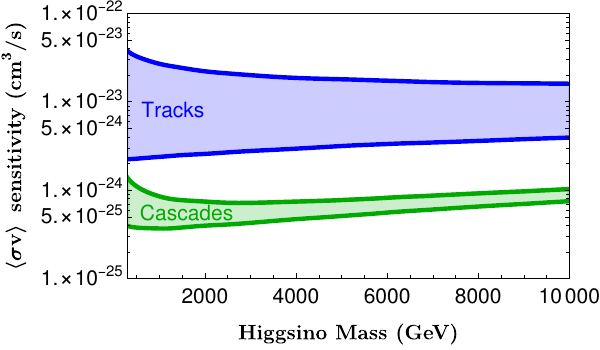}
    \caption{Sensitivity using tracks (blue) or cascade (green) neutrino signatures in IceCube for a search for Higgsino-like DM with some additional physics allowing it to compose the entire observed DM abundance. For tracks, only $\nu_\mu$ signal and background are possible, while for cascades only electron and tau neutrinos are included due to the large atmospheric $\nu_\mu$ background. Sensitivities are shown as bands based on a range of possible projections of background neutrino flux.}
    \label{fig:tracks_cascades}
\end{figure}

We optimize the size of the angular area of interest about the GC to minimize the required exposure for an exclusion (or detection), up to the limit of the detector's angular resolution. Fig.~\ref{fig:angular} shows the total exposure required for a 2$\sigma$ exclusion of thermally-produced Higgsino DM as a function of the radius of the cone of interest about the GC, using the optimistic background assumptions described above. As the DM signal is concentrated near the GC, smaller angular areas of observation about the GC lead to reduced exposure requirements, with a minimum at roughly 0.1\degree~when the signal begins to grow too faint. The vertical, dashed lines in Fig.~\ref{fig:angular} show the angular resolutions for track and cascade data for TeV-scale neutrinos in IceCube at 1\degree~and 10\degree~ respectively~\cite{IceCube:2019cia,IceCube:2023ame}, although for higher energy signals, we would expect better angular resolution. As we can see, the expected improvements in sensitivity from angular resolution enhancements are marginal for this analysis below a few degrees.

We now turn to the relative merits of different types of neutrino signals. while cascade events have lower angular resolution than track events, the effective area is larger and they provide sensitivity to more neutrino flavors, resulting in an improved reach for Higgsino-like DM relative to searches with track-like events, when searching for a signal from the GC. In Fig.~\ref{fig:tracks_cascades}, we compare the two sensitivities of the IceCube neutrino detector to track-like events (blue/upper shaded region) and cascade events (green/lower shaded region) from dark matter annihilations near the GC. Here we define the sensitivity band by the projections for the backgrounds discussed above. We see that the sensitivity of an analysis using cascades dominates over that for tracks, even with the most optimistic assumptions about the backgrounds.  In the remainder of this study, we therefore focus our analysis on cascade events.

\begin{figure}
    \centering
    \includegraphics[width=\imgscaling\columnwidth]{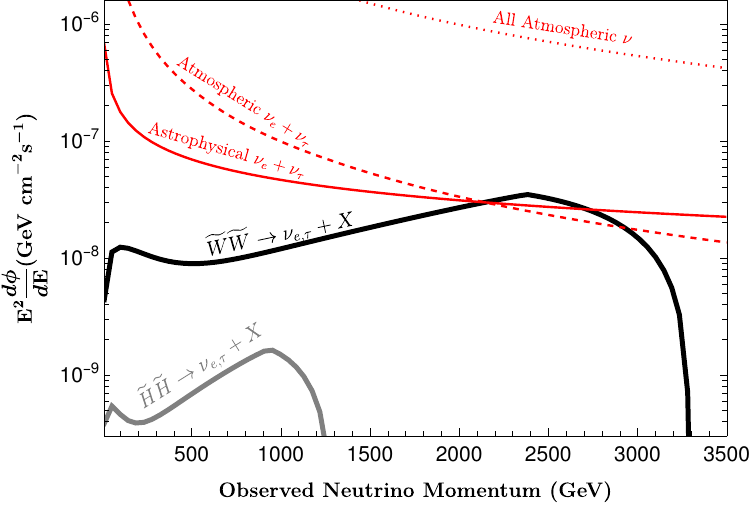}
    \caption{Neutrino flux for (black) Wino- and (gray) Higgsino-like neutralino DM shown as differential flux multiplied by the energy squared. Masses are set by the requirement that the neutralino produces the full DM relic abundance through thermal production, so the Wino-like neutralino mass is 2.8 TeV (black) and for Higgsino-like it is 1.1 TeV (gray). the $\nu_\mu$ signal is shown, but the spectrum is not visibly different for $\nu_{\mathrm{e}}$ and only very slightly suppressed for $\nu_\tau$. The astrophysical $\nu_\mu$ background (solid red) follows the power law described in Ref.~\cite{Silva:2023wol}, while atmospheric neutrinos (dashed red) follow the Honda model~\cite{Honda:2006qj}.}
    \label{fig:spectra}
\end{figure}

We consider the signal from angular area about the GC of 10\degree, as limited by the angular resolution for cascades in IceCube.  The resultant combined flux spectrum from Eq.~\ref{eq:sigFlux} for electron- and tau-neutrinos due to neutralino annihilation, normalized by a factor of $E^{2}$, is shown in Fig.~\ref{fig:spectra}.
We also show, in red, the atmospheric and astrophysical neutrino backgrounds described above.  We note that for both the DM signal and the astrophysical neutrino background, the spectrum is not visibly distinct for different neutrino flavors due to mixing over the distance traveled to Earth. In atmospheric neutrinos, however, there is a significant difference; muon-neutrinos dominate the spectrum, electron-neutrinos are subdominant, and the tau-neutrino contribution is negligible. 
While the astrophysical neutrino flux is not well characterized at energies as low as 1 TeV\footnote{The power law in description for astrophysical neutrinos Ref.~\cite{Silva:2023wol} is known to be unstable below 3 TeV.}, we assume the extrapolation is sufficiently accurate for our purposes. The signal flux is shown for both a Higgsino and Wino DM model, with DM masses 1.1 TeV and 2.8 TeV respectively to reproduce the observed DM density through thermal production.

\begin{figure*}
    \centering
    \includegraphics[width=\imgscaling\columnwidth]{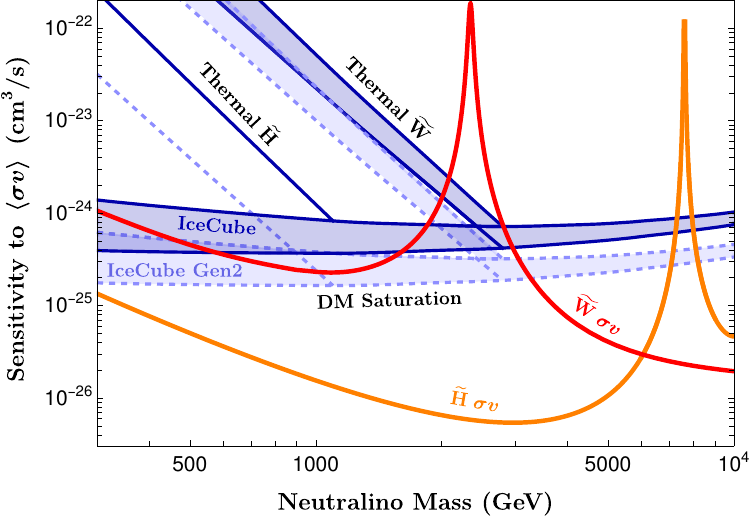}
    \includegraphics[width=\imgscaling\columnwidth]{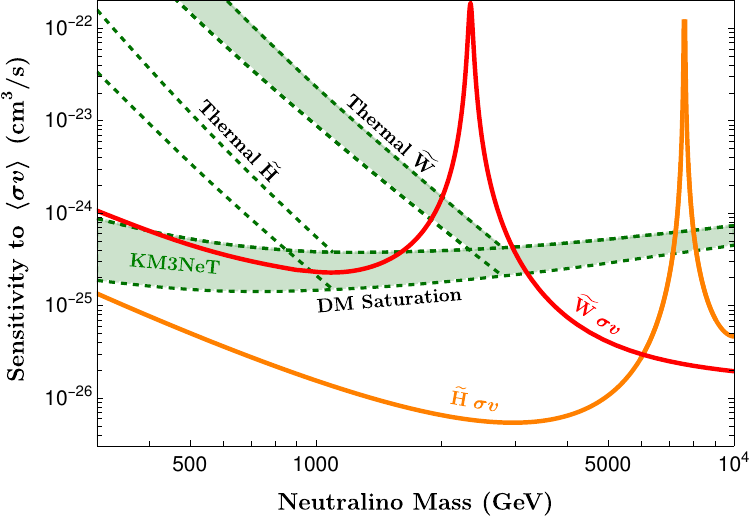}\\
    \vspace{6pt}
    \includegraphics[width=\imgscaling\columnwidth]{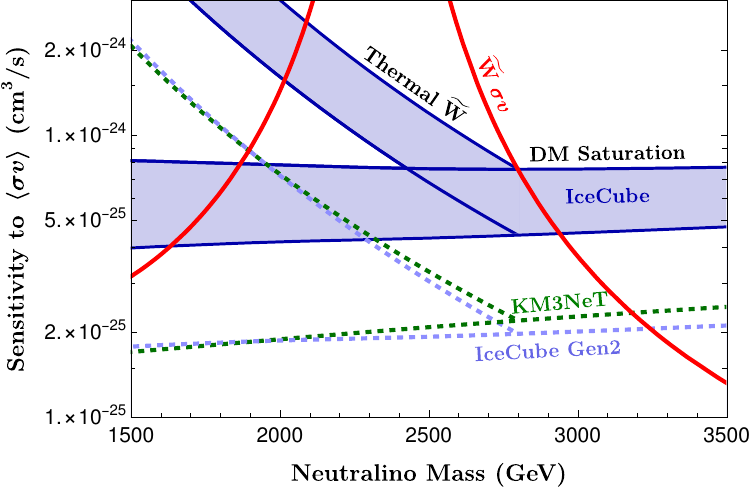}
    \caption{Comparison of IceCube (dark blue), IceCube gen2 (light blue), and KM3NeT (green) sensitivities to Neutralino DM. These are shown as IceCube (left), KM3NeT (right), and a focus on the region of interest for Wino exclusion with future experiments directly compared in the optimistic background assumption (below) The bands for each detector range from the most conservative case at the top, detecting only electron- and tau-neutrinos with both astrophysical and atmospheric backgrounds, to the most optimistic case at the bottom, detecting all neutrino flavors with only astrophysical neutrino backgrounds. The horizontal bands assume the neutralinos saturate the measured DM abundance while the diagonal bands are sensitivities assuming thermal production of Higgsinos (unfilled) and Winos (filled). The theoretically expected annihilation cross section is shown for Higgsinos (orange) and Winos (red).}
    \label{fig:sensitivity_bands}
\end{figure*}

\section{Results}
\label{sec:Res}
The overall current sensitivity for IceCube to Higgsino- and Wino-like DM is calculated assuming a 10 year run using cascade data for a total exposure $\sim 2\times 10^{11}$ cm$^2$s for neutrino energies near 1 TeV. IceCube Gen2 is taken to improve the effective area by roughly a factor of five between enhanced atmospheric veto and additional detectors~\cite{Ishihara:2019aao,Clark:2021fkg}. These are compared against the expected KM3NeT performance by adjusting the exposure based in Fig. 19 in Ref.~\cite{KM3Net:2016zxf}, or roughly 9000 cm$^2$ for neutrino energies near 1 TeV, and we use a 10 year run again for consistency (total exposure $\sim 3\times 10^{12}$ cm$^2$s). KM3NeT is taken to be able to use a 3\degree ~radius about the GC based on limits for shower angular resolution~\cite{KM3Net:2016zxf}, while both IceCube curves assume cascade data observing a 10\degree~radius area. Depending on how well a dedicated analysis is able to filter atmospheric neutrinos, we show the relative sensitivities for Higgsino- and Wino-like DM, as compared to the theoretically calculated cross section, in Fig.~\ref{fig:sensitivity_bands}. The bands represent the range from a conservative sensitivity projection where only electron-neutrinos and tau-neutrinos are detected with astrophysical and atmospheric backgrounds, to an optimistic one with only astrophysical neutrino backgrounds and all flavors detected with tracks and cascades. The sensitivity to Higgsinos and Winos are shown together on the plot as the sensitivity to Higgsinos is roughly 6\% more strict than Winos due to the different mix of Z- and W-bosons produced in annihilation, which is not a visible distinction.

If we assume thermal production of the lightest neutralino,
the density scales directly as $m_\chi^2$. This has two implications, first that for masses larger than 1.1 TeV for Higgsino-like and 2.8 TeV for Wino-like DM, we overclose the universe, and the second is that for lower masses, as the J-factor scales with $\rho^2$, the total sensitivity falls directly as  $m_\chi^4$ as compared to a similar model that saturates the observed DM abundance. In Fig.~\ref{fig:sensitivity_bands}, we show the sensitivities to thermal production with multicolored bands Wino-like DM, and for ease of reading with just the lines for the outer edges of the bands for Higgsino-like DM. For all detectors, the roughly horizontal bands assume some additional physics adjusts the relic abundance to saturate the bound observed of DM.

Comparing the IceCube sensitivities to the theoretically predicted cross sections shown in Fig.~\ref{fig:sensitivity_bands}, it is of particular note that the thermal DM wino appears entirely discoverable with current IceCube data. The bottom plot then shows a more focused study of this region of sensitivity with IceCube, IceCube Gen2, and KM3NeT shown together, though for the latter two only the optimistic background scenario is shown. We note that even Wino LSPs that do not saturate the DM abundance could be discoverable with masses near 2 TeV, and for lighter masses, near-future experiments can exclude Wino-like DM entirely if it saturates the DM abundance.

It is of note that uncertainties in the shape of the DM distribution and in the exact background spectrum can modify our prediction, although the former effect is somewhat ameliorated by the wide angular area of observation for the IceCube curves and the latter by the conservative background assumption. While Higgsino-like DM thermally produced to saturate the DM abundance is out of reach for currently planned experiments, IceCube Gen2 and KM3NeT will bring us within a couple orders of magnitude of sensitivity, implying sensitivity to this model may be a few generations away. IceCube general WIMP searches similar to this more specific one are ongoing~\cite{IceCube:2017rdn,ANTARES:2020leh, IceCube:2023ies, IceCube:2023orq}, and so far demonstrate a similar exclusion level as discussed here at roughly $\left<\sigma v\right> \lesssim 10^{-24} \text{cm}^3/$s. High energy gamma ray searches have demonstrated similar reach in the photon channel~\cite{Cohen_wino_2013,Dessert:2022evk}, complementing the results shown here.

\section{Conclusion}
\label{sec:Con}
The current exposure level from IceCube is sufficient to exclude thermally produced, Wino-like DM with a dedicated search. The particulars of the efficiency of excluding backgrounds from atmospheric muons and muon-neutrinos at 1 TeV remain as a topic for future study to refine this prediction, particularly for other neutralino masses where sensitivity is less definite, but current deep learning and muon coincidence techniques appear promising. Future experiments, including data from IceCube Gen2 and KM3NeT can also exclude the full regime of Wino-like DM that don't thermally over-produce DM, assuming additional physics enhances neutralino production to the observed DM abundance. For Higgsino-like DM, exclusion is still another generation of detector away outside of a region of parameter space with particularly strong Sommerfeld enhancement at roughly 7 TeV.

\vspace{\baselineskip}
\noindent
{\it Acknowledgments}: We would like to thank Minjin Jeong for helpful discussions on IceCube backgrounds and effective areas. The work of B.S.~is supported by the U.S. Department of Energy under Award No. DESC0009959. The work of P.S.~is supported in part by NSF PHY-2014075. C. Rott acknowledges support from NSF Grant No. PHY-2309967 and from the National Research Foundation of Korea (NRF) for the Basic Science Research Program NRF-2020R1A2C3008356. 

%

\end{document}